\documentstyle[12pt]{article}

\def\abcd{\alpha,\beta,\gamma,\delta}
\def\pas{h}  

\def \Plv{Pain\-le\-v\'e}
\def \CRAS{C.~R.~Acad.~Sc.~Paris}
\def \AnnENS{Ann.~\'Ec.~Norm.~}
 
\def \D {\hbox{d}}

\begin{document}

\pagestyle{plain} 

\begin{center}
 {\bf TOWARDS SECOND ORDER LAX PAIRS TO DISCRETE}

 {\bf PAINLEV\'E EQUATIONS OF FIRST DEGREE}
\end{center}

\vskip 0.5 truecm

{\bf Robert Conte\dag} and {\bf Micheline Musette\ddag}

\medskip

\dag
Service de physique de l'\'etat condens\'e,
CEA Saclay,
\hfill \break \indent
F-91191 Gif-sur-Yvette Cedex,
France
\medskip

\ddag
Dienst Theoretische Natuurkunde,
Vrije Universiteit Brussel,
\hfill \break \indent
B-1050 Brussel,
Belgique

\vskip 0.5 truecm

\vglue 1.0truecm
\baselineskip=12truept
\baselineskip=14truept
 

\vskip 0.8 truecm

\noindent {\it Keywords}
\par Painlev\'e equations
\par discrete Painlev\'e equations
\par Lax pairs
\par discrete Lax pairs
 
\vskip 0.8 truecm

{\it {\bf Abstract} -- 
We investigate the question of finding discrete Lax pairs for the six
discrete Painlev\'e equations (Pn).
The choice we make is to discretize the pairs of Garnier,
once converted to matricial form.
}

\vfill 

\noindent 
{\it Integrability and chaos in discrete systems},
Chaos, solitons and fractals {\bf } (1998),
eds.~I.~Antoniou and F.~Lambert.
Brussels, 2--6 July 1997. 
\hfill \break
\noindent

\rightline{\noindent 
           \hskip 1 truemm 22 May 1998
           \hskip 5 truemm S98/018 solv-int/9803013}

\eject

\tableofcontents \vfill \eject

\baselineskip=14truept


\section{Introduction}
\indent

In their programme of defining functions by nonlinear 
ordinary differential equations (ODEs),
Painlev\'e and Gambier built a list \cite{GambierThese} of fifty second order,
first degree such equations, the general solution of which is uniformizable;
among them, only six define new functions,
called the Painlev\'e transcendents (Pn).
There have been recently many achievements \cite{GNRCargese}
in obtaining some discrete analogues of these fifty equations.
However, at present time,
the nice features of the continuous case
(degree one in $u''$ for the ODEs,
order two for the Lax pairs of the six (Pn) equations,
coalescence cascade from (P6) to (P1) for the ODEs as well as for their 
Lax pairs)
are far from being all implemented in the discrete case.
Table \ref{tableExistingLax} outlines some of the gaps to be filled.

\begin{table}[h] 
\caption[garbage]{
Current state of the discrete (Pn) equations and their Lax pair.
Column 1 contains the type (difference or $q-$difference) of the discrete (Pn),
column 2 the degree of this discrete equation in $\overline{u}$,
column 3 the order of the Lax pair if one is known,
column 4 the appropriate references.
A blank denotes a missing information.
Ideally, there should be at least one entry for each d--(Pn),
only digits $1$ in column 2 and digits $2$ in column 3.
}
\vspace{0.2truecm}
\begin{center}
\begin{tabular}{| l | c | c | l |}
\hline
$
\displaystyle
\hbox{d--(Pn)}
\atop
\displaystyle
q-\hbox{(Pn)}
$
&
degree
&
order of Lax pair 
&
references
\\ \hline \hline 
 d--(P6) &    $2$    &    $ $    & \cite{RGO1998}
\\ \hline 
$q-$(P6) &    $2$    &    $2$    & \cite{JS1996}
\\ \hline 
 d--(P5) &    $ $    &    $ $    &
\\ \hline 
$q-$(P5) &    $1$    &    $ $    & \cite{RGH1991}
\\ \hline 
 d--(P4) &    $1$    &    $ $    & \cite{RGH1991}
\\ \hline 
 d--(P3) &    $2$    &    $4$    & \cite{GNPRS1994}
\\ \hline 
$q-$(P3) &    $1$    &    $4$    & \cite{RGH1991,PNGR1992}
\\ \hline 
 d--(P2) &    $1$    &    $2$    & \cite{GNPRS1994}
\\ \hline 
 d--(P1) &    $1$    &    $2$    & \cite{FGR}
\\ \hline 
 d--$\wp$&    $1$    &    $2$    & \cite{Potts1987}
\\ \hline \hline 
\end{tabular}
\end{center}
\label{tableExistingLax}
\end{table}

In this paper,
we address the question of finding a second order Lax pair for each discrete
(Pn) equation.

In section \ref{PnGarnier},
we present the five (Pn) equations chosen by Garnier,
which have some advantages over the six usual ones.
In section \ref{sectionVariousLax},
we review the different kinds of continuous Lax pairs
and discuss them according to their relevance for discretization.
In section \ref{sectionGarnierMatricialPairs},
we convert the scalar ``Lax'' pairs given by Garnier in 1911
into an equivalent traceless matricial form,
so as to remove the
apparent singularity unavoidable in the scalar form.
In section \ref{sectionDiscretizationRules},
we recall the discretization rules which we previously established.
In section \ref{sectionApplication},
we give some preliminary results.

\section{The five (Pn) equations of Garnier and their coalescence}
\label{PnGarnier}
\indent

The (P6) equation for $u(x)$
possesses in the plane of $u$ four poles $(\infty,0,1,x)$ with the same residue
equal to $1/2$,
and it depends on four parameters  $\abcd$.
The scheme of the successive coalescences of these four poles
\begin{equation} 
\matrix{
{ }                  &{ }     &{}             &{ }       &{(1/2,3/2)}
&{} &{} &\cr
{ }                  &{}      &{ }            &{\nearrow}&{ }          
&{\searrow}&{ }&\cr
{(1/2,1/2,1/2,1/2)}&{\rightarrow}&{(1/2,1,1/2)}&{ }       &{}           
&{ }       &{(2)}       &\cr
{ }                  &{}      &{ }            &{\searrow}&{ }          
&{\nearrow}&{ }       &\cr
{ }                  &{ }     &{}             &{ }       &{(1,1)}    
&{ }       &{} &\cr
} 
\end{equation}
defines,
from (P6)$(u,x,\alpha,\beta,\gamma,\delta)$, 
four other equations with four parameters,
chosen as follows by Garnier \cite{GarnierThese} 
\label{P6P0def}
\begin{eqnarray*}
({\rm P6})\
u''
&=&
{1 \over 2} \left[{1 \over u} + {1 \over u-1} + {1 \over u-x} \right] u'^2
- \left[{1 \over x} + {1 \over x-1} + {1 \over u-x} \right] u'
\\
& &
+ {u (u-1) (u-x) \over x^2 (x-1)^2} 
  \left[\alpha + \beta {x \over u^2} + \gamma {x-1 \over (u-1)^2} 
        + \delta {x (x-1) \over (u-x)^2} \right],
\\
({\rm P5})\ 
u''
&=&
\left[{1 \over 2 u} + {1 \over u-1} \right] u'^2
- {u' \over x}
+ {(u-1)^2 \over x^2} \left[ \alpha u + {\beta \over u} \right]
+ \gamma {u \over x}
+ \delta {u(u+1) \over u-1},
\nonumber
\\
({\rm P4'})\ 
u''
&=&
{u'^2 \over 2 u} + \gamma \left({3 \over 2} u^3 + 4 x u^2 + 2 x^2 u\right)
+ 4 \delta (u^2 + x u) - 2 \alpha u + {\beta \over u}
\\
({\rm P3'})\ 
u''
&=&
{u'^2 \over u} - {u' \over x} + {\alpha u^2 + \gamma u^3 \over 4 x^2}
 + {\beta \over 4 x}
 + {\delta \over 4 u}
\\
({\rm P2'})\ 
u''
&=&
\delta (2 u^3 + x u) + \gamma (6 u^2 + x) + \beta u + \alpha
\\
({\rm J'})\ 
u''
&=&
2 \delta u^3 + 6 \gamma u^2 + \beta u + \alpha.
\end{eqnarray*}
Under the group of homographic transformations on $u$,
the (P4') equation is equivalent to  (P4) for $\gamma \not=0$;
the (P3') equation is that defined by Painlev\'e
(Ref.~\cite{PaiCRAS1906} p.~1115) in the class of (P3),
the (P2') belongs to the class of (P2) for $\delta \not=0$
and to that of (P1) for $\delta=0,\gamma \not=0$.
This list has been terminated by the elliptic equation 
which is the autonomous limit of (P2'), which we denote (J') (like ``Jacobi'').

This is the most advantageous list of equations to be considered
for the study of Lax pairs,
for it allows to take full advantage of the four parameters $\abcd$
appearing in each equation.

The successive coalescences 
\begin{equation} 
\matrix{
{}    &{}           &{}    &{}        &{\rm (P4')}&{}        &{}       &{} &{}
 &\cr
{}    &{}           &{}    &{\nearrow}&{}     &{\searrow}&{}       &{} &{} &\cr
{\rm (P6)}&{\rightarrow}&{\rm (P5)}&{}        &{}     &{}        &{\rm (P2')}
&{\rightarrow} 
&{\rm(J')} &\cr
{}    &{}           &{}    &{\searrow}&{}     &{\nearrow}&{}       &{} &{} &\cr
{}    &{}           &{}    &{}        &{\rm (P3')}&{}    &{}       &{} &{} &\cr
} 
\end{equation}
from an equation 
$E(x,u,\abcd)=0$ 
to another equation 
$E(X,U,A,B,C,D)=0$
are described by the following Poincar\'e perturbations \hfill \break
$(x,u,\abcd) \to (X,U,A,B,C,D,\varepsilon)$ 
\begin{eqnarray*}
6 \to 5:
& &
(x,u,\alpha,\beta,\gamma,\delta)
=
(1+\varepsilon X,U,
A,B,\varepsilon^{-1}C-\varepsilon^{-2}D,\varepsilon^{-2}D),
\\
5 \to 4':
& &
(x,u,\alpha,\beta,\gamma,\delta)
=
(1+ \varepsilon X, \varepsilon U/2,
 2 C \varepsilon^{-4} + 28 D \varepsilon^{-3}, B/4,
\\
& &
\phantom{(x,u,\alpha,\beta,\gamma,\delta)=}
-4 C \varepsilon^{-4} - 60 D \varepsilon^{-3},
 2 A \varepsilon^{-2} -  2 C \varepsilon^{-4}- 32 D \varepsilon^{-3}),
\\
5 \to 3':
& &
(x,u,\alpha,\beta,\gamma,\delta)
=
(X,1 + \varepsilon U,
\varepsilon^{-1} A/4 + \varepsilon^{-2} C/8,
- \varepsilon^{-2} C/8, \varepsilon B/4, \varepsilon^{2} D/8),
\\
4' \to 2' :
& &
(x,u,\alpha,\beta,\gamma,\delta)
=
(\varepsilon^{2} X/4,
1 + \varepsilon U,
 -4 B \varepsilon^{-4} +96 C \varepsilon^{-5}- 24 D \varepsilon^{-6},
\\
& &
\phantom{(x,u,\alpha,\beta,\gamma,\delta)=}
 16 A \varepsilon^{-3} - 8 B \varepsilon^{-4} +48 C \varepsilon^{-5}
- 8 D \varepsilon^{-6},
\\
& &
\phantom{(x,u,\alpha,\beta,\gamma,\delta)=}
-32 C \varepsilon^{-5} +16 D \varepsilon^{-6},
 48 C \varepsilon^{-5}- 16 D \varepsilon^{-6}),
\\
3' \to 2' :
& &
(x,u,\alpha,\beta,\gamma,\delta)
=
(1+\varepsilon^2 X/2,1 + \varepsilon U,
16 B \varepsilon^{-4} - 64 C \varepsilon^{-5} - 32 D \varepsilon^{-6},
\\
& &
\phantom{(x,u,\alpha,\beta,\gamma,\delta)=}
32 D \varepsilon^{-6},
-8 B \varepsilon^{-4} + 48 C \varepsilon^{-5} + 16 D \varepsilon^{-6},
\\
& &
\phantom{(x,u,\alpha,\beta,\gamma,\delta)=}
16 A \varepsilon^{-3} -  8 B \varepsilon^{-4} + 16 C \varepsilon^{-5}
 - 16 D \varepsilon^{-6}),
\\
2' \to {\rm J'}:
& &
(x,u,\alpha,\beta,\gamma,\delta)
=
(\varepsilon X, \varepsilon^{-1} U, 
\varepsilon^{-3} A, \varepsilon^{-2} B, \varepsilon^{-1} C, D)
\end{eqnarray*}
in which $\varepsilon$ goes to zero.
These six transformation laws on $x$ and $u$ are affine,
contrary to those of the usual coalescence cascade.

\section{Various Lax pairs of (Pn) in the continuous case}
\label{sectionVariousLax}
\indent

Lax pairs can be obtained either by the isomonodromic deformation of
a given second order linear differential equation
(defined either in scalar form or in matricial form),
or by the reduction of the Lax pair for some partial differential equation
(PDE) \cite{FlaschkaNewell},
which makes three possibilities.

\subsection{Lax pairs by scalar isomonodromy (Garnier)}
\label{GarnierPairs}
\indent

In order to find (P6),
Richard Fuchs had to assume in the second order scalar equation 
\begin{eqnarray}
& &
\partial_t^2 \psi + (S/2) \psi=0,\
\end{eqnarray}
the presence
of four Fuchsian singularities, put {\it a priori} at 
the points $t=\infty,0,1,x$.
As shown by Poincar\'e (Ref.~\cite{Poincare1883} pp.~217--20),
the maximum number of apparent singularities
(a singularity is apparent iff in its neighborhood the ratio
of any two solutions $\psi$ is singlevalued)
which can be added in this case (second order, four Fuchsian singularities)
is one.
If no apparent singularity is added,
a case which corresponds to the Heun equation \cite{Heun1889,HeunWorkshop},
the monodromy is trivial.
By adding one apparent singularity, at a location denoted $t=u$,
R.~Fuchs \cite{Fuchs1905,Fuchs1907} found that $u(x)$ had to satisfy the (P6) 
equation.
This result has been completed by Garnier,
who established a coalescence cascade starting from (P6)
and yielding scalar pairs for (P5), (P4'), (P3'), (P2') and (J')
while respecting a remarkable symmetry between $t$, $x$ and $u$.

These scalar Lax pairs of Garnier are linear in $\abcd$ and 
characterized by the two 
quantities $(S,C)$,
with the cross-derivative condition $Z=0$
\begin{eqnarray}
& &
\partial_t^2 \psi + (S/2) \psi=0,\
\label{eqGarnierScalarODE}
\\
& &
\partial_x \psi + C \partial_t \psi -(1/2) C_t \psi = 0,\
\label{eqGarnierScalarPDE}
\\
& &
Z \equiv S_x + C_{ttt} + C S_t + 2 C_t S=0.
\end{eqnarray}
The value of $S$ is given by
\begin{eqnarray}
& &
-{S \over 2}=
{3/4 \over (t-u)^2}
+ {\beta_1 u' + \beta_0 \over (t-u) e_1}
+{[(\beta_1 u')^2 - \beta_0^2] e_0 + f_{\rm G}(u) \over e_2}
+ f_{\rm G}(t),
\label{eqS}
\end{eqnarray}
in which the function $f_{\rm G}$ and the various scalars are defined 
in Table \ref{tableGarnierScalaire}.

\tabcolsep= 0.5 truemm

\begin{table}[h] 
\caption[garbage]{
Scalar Lax pairs of the Painlev\'e equations
}
\vspace{0.2truecm}
\begin{center}
\begin{tabular}{| l | c | c | c | c | c |}
\hline
&
(P6)
&
(P5)
&
(P4')
&
(P3')
&
(P2')
\\ \hline \hline 
$
\beta_1
$
&
$
\displaystyle
-{x (x-1) \over 2 (u-x)}
$
&
$
\displaystyle
-{x \over 2 (u-1)}
$
&
$
1/4
$
&
$
\displaystyle
-{x \over 2 u}
$
&
$
-1
$
\\ \hline 
$
\beta_0
$
&
$
-u+1/2
$
&
$
-u+1/2
$
&
$
1/2
$
&
$
-1/2
$
&
$
0
$
\\ \hline 
$
f_G(z)
$
&
$
\displaystyle
{a \over z^2} + {b \over (z-1)^2}
\atop
\displaystyle
+{c \over (z-x)^2} + {d \over z (z-1)}
$
&
$
\displaystyle
{a \over z^2} - {b x \over (z-1)^3} 
\atop
\displaystyle
+ {c x^2 \over (z-1)^4} + {d \over (z-1)^2}
$
&
$
\displaystyle
- {a \over z^2} 
- {\delta (z+2x) \over 4}
\atop
\displaystyle
-b 
- {\gamma (z+2x)^2 \over 16} 
$
&
$
\displaystyle
a - {b x \over z^3}
\atop
\displaystyle
+{c x^2 \over z^4} + {d \over z}
$
&
$
\displaystyle
2 \alpha z 
+ \beta z^2
\atop
{\displaystyle
+ 2 \gamma (2 z^3 + x z) 
\atop
\displaystyle
+ \delta (z^4 + x z^2)
}
$
\\ \hline 
$
\alpha
$
&
$
2(a+b+c+d+1)
$
&
$
2(a+d+1)
$
&
$
-4 b
$
&
$
8 d
$
&
$
$
\\ \hline 
$
\beta
$
&
$
-2(a+1/4)
$
&
$
-2(a+1/4)
$
&
$
-8 a -2
$
&
$
8 b
$
&
$
$
\\ \hline 
$
\gamma
$
&
$
2(b+1/4)
$
&
$
2 b
$
&
$
$
&
$
16 a
$
&
$
$
\\ \hline 
$
\delta
$
&
$
-2 c
$
&
$
-2 c
$
&
$
$
&
$
-16 c
$
&
$
$
\\ \hline 
$
e_0
$
&
$
\displaystyle
{u-x \over u (u-1)}
$
&
$
1/u
$
&
$
1/u
$
&
$
1
$
&
$
1
$
\\ \hline 
$
e_1
$
&
$
t(t-1)
$
&
$
t(t-1)
$
&
$
-t
$
&
$
t
$
&
$
1
$
\\ \hline 
$
e_2
$
&
$
t(t-1)(t-x)
$
&
$
t(t-1)^2
$
&
$
t
$
&
$
t^2
$
&
$
1
$
\\ \hline 
$
- C
$
&
$
\displaystyle
{t(t-1) (u-x) \over (t-u) x(x-1)}
$
&
$
\displaystyle
{t(t-1) (u-1) \over (t-u) x}
$
&
$
\displaystyle
2 { t  \over t-u}
$
&
$
\displaystyle
{t u \over (t-u) x}
$
&
$
\displaystyle
{1/2 \over t-u}
$
\\ \hline \hline 
\end{tabular}
\end{center}
\label{tableGarnierScalaire}
\end{table}

\subsection{Lax pairs by matricial isomonodromy (Jimbo and Miwa)}
\indent

Equivalently, the second order deformation equation can be replaced
by a first order, two-component, matricial system.
The advantage is the possibility to get rid of the apparent singularity,
unavoidable in the scalar case,
by a change of basis.
The four Fuchsian singularities are similarly put at the {\it a priori} 
locations $t=\infty,0,1,x$.
The resulting matricial Lax pairs obtained by Jimbo and Miwa \cite{JimboMiwaII}
are different from those of Garnier once converted to matricial form;
in particular, they are no more linear in $\abcd$, but algebraic.
Some of them, namely those for (P3), (P5) and (P6),
are restricted by nonvanishing conditions on the parameters.
We give here traceless Lax pairs of (P3), (P5), (P6) 
for generic values of $\abcd$
(indeed, in Ref.~\cite{JimboMiwaII}, some of these parameters are restricted
to numerical values, and the Lax pair of (P6) is not traceless).
In the following formulae,
the symbols $\sigma_k$ denote the Pauli matrices
\begin{eqnarray}
& &
\sigma_1=\pmatrix{0 & 1 \cr 1 & 0 \cr },\ 
\sigma_2=\pmatrix{0 &-i \cr i & 0 \cr },\ 
\sigma_3=\pmatrix{1 & 0 \cr 0 &-1 \cr },\ 
\sigma_j \sigma_k= \delta_{jk} + i \varepsilon_{jkl} \sigma_l,\
\label{eqPauli}
\\
& &
\sigma^{+}=\pmatrix{0 & 1 \cr 0 & 0 \cr },\ 
\sigma^{-}=\pmatrix{0 & 0 \cr 1 & 0 \cr },
\nonumber
\end{eqnarray}
while the symbol $Z$ is the commutator of the Lax pair
\begin{eqnarray}
& &
\partial_x \psi = L \psi,\
\partial_t \psi = M \psi,\
Z = [\partial_x - L, \partial_t - M]
\end{eqnarray}
and the symbol $E$ denotes the (Pn) equation.

The pair of (P3) \cite{JimboMiwaII} for $(\gamma,\delta) \not=(0,0)$ is
\begin{eqnarray}
L&= &
{t \over x}\bigg[
(d x/4 + t^{-2} (c x /4 - z)) \sigma_3
\nonumber
\\
& &
+ (t^{-2} - t^{-1}u)              w      \sigma^{+}
+ (t^{-1} V - t^{-2} z (z-c x/2)) w^{-1} \sigma^{-}
\bigg]
\label{eqLaxP3JMpartiex}
\\
M&= &
(d x/4 - t^{-2} (c x /4 - z) - t^{-1} (\beta d^{-1} -2)/4) \sigma_3
\nonumber
\\
& &
- (t^{-2} + t^{-1} u)             w      \sigma^{+}
+ (t^{-1} V + t^{-2} z (z-c x/2)) w^{-1} \sigma^{-}
\label{eqLaxP3JMpartiet}
\\
u^2 Z &= &
((1/8) t^{-2}
 (2 x \sigma_3 +(x^2 u^{-2}(u'-d)+(\beta d^{-1} -1) x u^{-1})w^{-1} \sigma^{-})
\nonumber
\\
& &
-(1/8) t^{-1} (x^2 u^{-2}(u'-d)+ x) w^{-1} \sigma^{-}) E
\\
4 z &= &
 x u' / u^2 + c x - d x/u^2 + (\beta d^{-1} -1)/u,\
\\
V &= &
- z^2 u + z (c x u + \beta d^{-1} -2)/2 + (\alpha + 2 c - \beta c d^{-1}) x/8,\
\\
w'/w &= &
-u'/u + d/u + (1 -(1/2) \beta d^{-1}) /x 
\\
c^2 & = & \gamma,\
d^2=- \delta.
\end{eqnarray}

The pair of (P5) \cite{JimboMiwaII} for $\delta \not= 0$ is
\begin{eqnarray}
L&= &
d t \sigma_3/2
\nonumber
\\
& &
+ x^{-1} 
 ( z + \theta_0 -u     (z+(\theta_0 - \theta_1 + \Theta_{\infty})/2))w 
\sigma^{+}
\\
& &
+ x^{-1} 
 (-z            +u^{-1}(z+(\theta_0 - \theta_1 + \Theta_{\infty})/2))w^{-1}
\sigma^{-}
\\
M&= &
(d {x \over 2} + {z + \theta_0/2 \over t}
   - {z + (\theta_0 + \Theta_{\infty})/2 \over t-1}) \sigma_3
\nonumber
\\
& &
+
((z + \theta_0)/t - u (z + (\theta_0 - \theta_1 + \Theta_{\infty})/2)/(t-1))w
\sigma^{+}
\nonumber
\\
& &
+ (-z/t + u^{-1} (z + (\theta_0 + \theta_1 + \Theta_{\infty})/2)/(t-1))w^{-1}
\sigma^{-}
\\
Z&= &
\pmatrix{-u & u(t-1-tu)w \cr (t+u-tu)/w & u \cr} {x \over t (t-1)u (u-1)}E
\\
z&= &
-x{u'- d u \over 2 (u-1)^2}
+{(3 \theta_0+\theta_1+\Theta_{\infty})-(\theta_0-\theta_1+\Theta_{\infty})u
  \over 4 (u-1)}
\\
w'/w&= &
[-2z - \theta_0 + u (z + (\theta_0 - \theta_1 + \Theta_{\infty})/2)
           + u^{-1} (z + (\theta_0 + \theta_1 + \Theta_{\infty})/2)]/x
\\
d^2 &=& -2 \delta,\
(\theta_0 - \theta_1 + \Theta_{\infty})^2 = 8 \alpha,\
(\theta_0 - \theta_1 - \Theta_{\infty})^2 = - 8 \beta,\
(1 - \theta_0 - \theta_1)= \gamma/d.
\nonumber 
\end{eqnarray}

The traceless pair of (P6) for
$(2 \alpha-1,2 \beta+1,2 \gamma-1, \delta)\not=(0,0,0,0)$
can be found in Ref.~\cite{CargeseMahoux}.

\subsection{Lax pairs by reduction of a PDE}
\indent

One first selects a PDE which admits a second order Lax pair.
The reductions create singularities at four points at most,
which must then be moved by homography to $t=\infty,0,1,x$.
The resulting Lax pairs are generically algebraic in $\abcd$,
maybe with some restrictions on them.

One such system is the pumped Maxwell-Bloch system,
defined either in the complex form
\cite{BZM1987}
\begin{eqnarray}
& &
e_T=\rho,\
\rho_X= N e,\
\overline{e}_T=\overline{\rho},\
\overline{\rho}_X= N \overline{e},\
N_X+(\rho \overline{e} + \overline{\rho} e)/2 - 4 s=0,
\label{eqMaxwellBlochComplex}
\end{eqnarray}
with $s$ constant (the system is ``pumped'' when $s$ is nonzero),
or in the real form with one more arbitrary function $m_1(T)$
\begin{eqnarray}
& &
e_T=\rho,\
\rho_X= N e +m_1(T),\
N_X+ \rho e - 4 s=0.
\label{eqMaxwellBlochReal}
\end{eqnarray}
One does not know a possible common parent to these two systems.
These systems describe phenomena in nonlinear optics,
stimulated Raman scattering
and self-induced transparency.
Their respective Lax pairs are,
for the complex system (\ref{eqMaxwellBlochComplex})
\cite{KRT}
\begin{eqnarray}
L &=& {1 \over 2}  \pmatrix{0 & e \cr - \overline{e} & 0 \cr}
               + f \pmatrix{1 & 0 \cr 0 & -1 \cr},\
M = {1 \over 4 f}\pmatrix{N & - \rho \cr - \overline{\rho} & -N \cr},\ 
\label{eqLaxMaxwellBlochComplex}
\end{eqnarray}
for the real system (\ref{eqMaxwellBlochReal})
\cite{ClarksonMansfieldMilne} 
\begin{eqnarray}
{\hskip -10.0 truemm}
L &=& {1 \over 2}  \pmatrix{0 & e \cr - e & 0 \cr}
               + f \pmatrix{1 & 0 \cr 0 & -1 \cr},
M = {1 \over 4 f}
    \pmatrix{N & - \rho + m_1/(2 f) \cr - \rho - m_1/(2 f) & -N \cr}
\label{eqLaxMaxwellBlochReal}
\end{eqnarray}
with $f^2=2 s T + \lambda^2$
and $\lambda$ an arbitrary complex constant.

A Lax pair of (P3)
is obtained by the following reduction of the pumped
real Maxwell-Bloch system
(\ref{eqMaxwellBlochReal})
\cite{WinternitzSainteAdele,MilneThesis,CMMontreal}
\begin{eqnarray}
x & = & X T^{1/2},\
e=- i T^{1/2} \big[u'/u + c u + d/u \big],\
m_1 = - i (\alpha d - \beta c),\
s=- c d/2,\
\\
t & = & T^{-1/2} f(T),\
\\
L &= &
(1/2)(u'/u + c u + d/u) \sigma_3 + t \sigma_1,\
\label{eqLaxP3partiex}
\\
M &= & \Big[
(x/t) L +t^{-2} c d x \sigma_1/2 +t^{-3} (-\alpha d + \beta c) \sigma_3 /4
\label{eqLaxP3partiet}
\\
& &
-t^{-2} (\alpha u + \gamma x u^2 + c (x u)') \sigma^{+}/2
+t^{-2} (\beta /u - \delta x/u^2 - d (x/u)') \sigma^{-}/2
\Big] {t^2 \over t^2 + c d},
\nonumber
\\
Z &= &
{x \over 2 t u} (\sigma_3 - t^{-1} (c u \sigma^{+} + d u^{-1} \sigma^{-})) 
{t^2 \over t^2 + c d} E,
\\
c^2 & = & \gamma, d^2=- \delta.
\end{eqnarray}
The transition matrix
$ \pmatrix{t^{1/2} & 0 \cr 0 & t^{-1/2} \cr}$
makes this pair invariant by parity on $t$.
For $\gamma \delta=0$,
it has two nonFuchsian singularities $t^2=\infty,0$.
For $\gamma \delta\not=0$,
it has three singularities in the variable $-t^2/(c d)$~:
two Fuchsian $0$ and $1$,
one nonFuchsian at infinity,
i.e.~one more than expected for (P3),
and there exists no coalescence of two of these three singularities which
preserves (P3).
Therefore this is in fact a pair of (P5) for $\delta=0$ 
\cite{WinternitzSainteAdele}
since (P5) for $\delta=0,\gamma\not=0$ is equivalent to the full (P3) under
a birational transformation \cite{Gromak1975,FokasAblowitz1982}.
\medskip

Another Lax pair of (P3) is obtained by a reduction of the unpumped complex
Maxwell-Bloch system 
(\ref{eqMaxwellBlochComplex}),
but this one is restricted to $ (\gamma,\delta)\not=(0,0)$
\begin{eqnarray}
x & = & X T^{\theta},\
e            = T^{(1-2 \theta_k) \theta},\
\overline{e} = T^{(1+2 \theta_k) \theta},\
s=0,\
t= T^{- \theta} f(T),\
\\
L &= &
-(1/2)(u'/u + d/u +1/x) \sigma_1 
+(\alpha/(c x) + c u) (\sigma_3 + \sigma_2/i)/2 
-t \nu x /c,
\\
M &= & \Big[
t^{-2} c d \nu^{-1} u^{-1} (u'/u + d/u - (\beta + d)/(d x)) + t^{-1} \alpha/c
\Big] (\sigma_3 + \sigma_2/i)/4
\\
& &
+ \Big[- \nu x^2/c \sigma_3
+ t^{-1} (\alpha/c) \sigma_2/i
-t^{-1} x (u'/u + d/u + 1/x) \sigma_1
-t^{-2} \nu^{-1} c^2 d \sigma_1
\Big]/4
\nonumber
\\
Z &= &
(1/4) u^{-2}
[t^{-2} \nu^{-1} (\sigma_3 + \sigma_2/i) - t^{-1} x u \sigma_1] E,
\\
\theta_k & = & - \alpha /(4 c)
\\
c^2 & = & \gamma, d^2=- \delta.
\end{eqnarray}
in which $\theta$ is arbitrary.
The arbitrary parameter $\nu$ just reflects the scaling law of (P3)
and it can be set for instance to $c$, $d$ or $1$.
This pair 
is much simpler than the one of Jimbo and Miwa
(\ref{eqLaxP3JMpartiex})--(\ref{eqLaxP3JMpartiet}),
and it possesses two nonFuchsian singularities $t=\infty,0$.
\medskip

A Lax pair of the full (P5) is obtained by a reduction of the pumped complex
system
(\ref{eqMaxwellBlochComplex})
\cite{KRT}
with the restriction $\delta\not=0$
\begin{eqnarray}
& & x = X T^{1/2},\
e            =T^{1/2 - t_j} F,\
\overline{e} =T^{1/2 + t_j} G,\
s=-\delta/16,\
\\
& & t=2 T^{-1/2} f(T) /d +1/2
\\
& & L= \pmatrix{(2 t -1) 4 d & F/2 \cr -G/2 & -(2 t-1) 4 d \cr}
\\
& & M= {C_0 \over t} + {C_1 \over t-1} + (d/2) x \sigma_3
\\
& & Z= { d^2 x \over 8 (t-1) (u-1)^5}
 [\sigma_3 + u^{-1/2} ((t+u-t u) v \sigma^{+} + (t-1-t u) v^{-1} \sigma^{-})]E,
\nonumber
\\
& & F= u^{-1/2}\left[{u' + d u \over u-1} +2(t_c (u-1) - t_j u)/x\right] v
\\
& & G= u^{-1/2}\left[{u' + d u \over u-1} -2(t_c (u-1) - t_j u)/x\right] v^{-1}
\\
& & v'/v = (t_j (u^2 + 1) - t_c (u^2-1))/(x u)
\\
& & C_0+C_1 = - t_j \sigma_3 - (x/2) \pmatrix{0 & F \cr -G & 0 \cr}
\\
& & C_0-C_1 =
 \left[{x (u' + d u) \over (u-1)^2}-{(\gamma+d) (u+1) \over 2 d (u-1)}\right]
 \sigma_3
\nonumber
\\
& & 
\phantom{C_0-C_1 =}
+u^{-1/2} ((p+q) v \sigma^{+} + (p-q) v^{-1} \sigma^{-})
\\
& & 
p=-{x (u+1) (u' + d u) \over 2 (u-1)^2} u' +{(\gamma+d) u \over d (u-1)},\
q=t_c (u+1) - t_j u
\\
& & 
d^2=-2 \delta,\
t_c^2=- \beta/2,\
(t_c-t_j)^2=\alpha/2.
\end{eqnarray}

\vfill \eject
\subsection{Comparison of these three kinds of Lax pairs}
\indent

Our goal is to discretize as easily as possible without introducing
restrictions on $\abcd$.

The three kinds of Lax pairs are compared in
Table \ref{tableComparison}
according to several criteria.

\begin{table}[h] 
\caption[garbage]{
Advantages and inconveniences of the three kinds of Lax pairs.
The ideal situation would be that all entries be ``yes''.
}
\vspace{0.2truecm}
\begin{center}
\begin{tabular}{| l | c | c | c |}
\hline
&
scalar Garnier
&
matricial Jimbo and Miwa
&
by reduction
\\ \hline \hline 
no apparent singularity
&
no
&
yes
&
yes
\\ \hline 
linearity in $\abcd$
&
yes
&
no
&
no
\\ \hline 
easy confluence
&
yes
&
no
&
yes
\\ \hline 
arbitrary $\abcd$
&
yes
&
no
&
no
\\ \hline \hline 
\end{tabular}
\end{center}
\label{tableComparison}
\end{table}

We therefore choose the scalar pairs of Garnier
and, before attempting to discretize,
we remove the apparent singularity by converting them to matricial form.

{\it Remark}.
If one first discretizes a PDE then performs a discrete reduction,
the resulting discrete ODE is not explicit but is defined by the system
made of the discrete PDE and the discrete nonlinear relation defining the 
reduction; for details, see section 2.4 in Ref.~\cite{GNRCargese}.


\section{The Garnier matricial Lax pairs}
\label{sectionGarnierMatricialPairs}
\indent

From the scalar Garnier pairs
(\ref{eqGarnierScalarODE})--(\ref{eqGarnierScalarPDE}),
let us build an equivalent matricial pair which does not contain any more
the apparent singularity $t=u$.

Let us choose the two matrices as traceless
\begin{eqnarray}
& &
\left(\partial_x - \pmatrix{L_{11} & L_{12} \cr L_{21} & -L_{11} \cr}\right)
\pmatrix{\psi_1 \cr \psi_2 \cr}=0,\
\\
& &
\left(\partial_t - \pmatrix{M_{11} & M_{12} \cr M_{21} & -M_{11} \cr}\right)
\pmatrix{\psi_1 \cr \psi_2 \cr}=0.
\end{eqnarray}
The elimination of one component, say $\psi_1$,
defines the scalar system
\begin{eqnarray}
& &
\left(\partial_t^2 - {M_{21,t} \over M_{21}} \partial_t 
+ M_{11,t} - M_{11}^2 - M_{11} {M_{21,t} \over M_{21}} - M_{12} M_{21}
\right)
\psi_2=0,\
\label{eqConvertedMatrixODE}
\\
& &
\left(\partial_x - {L_{21} \over M_{21}} \partial_t 
+ L_{11} - M_{11} {L_{21} \over M_{21}}
\right)
\psi_2=0.
\label{eqConvertedMatrixPDE}
\end{eqnarray}
The coefficients of this scalar Lax pair have simple poles at $M_{21}=0$.
Let us first convert the double pole $t=u$ of the coefficients
of the pair of Garnier into a simple pole,
by the transformation
\begin{eqnarray}
& &
\Psi= \sqrt{t-u} \psi,\
\\
& &
\left(\partial_t^2 - {1 \over t-u} \partial_t
 + \left({S \over 2} + {3/4 \over (t-u)^2} \right)
\right) \Psi=0,
\label{eqConvertedGarnierODE}
\\
& &
\left(\partial_x+C \partial_t+{u' -(C + (t-u) C_t) \over 2 (t-u)}\right)\Psi=0,
\label{eqConvertedGarnierPDE}
\end{eqnarray}
with $S$ given by (\ref{eqS}) and $C$ by Table \ref{tableGarnierScalaire}.
The six coefficients $(L_{jk},M_{jk})$ of the Lax pair are obtained by
identifying 
the singularities $M_{21}=0$ and $t-u=0$ in
(\ref{eqConvertedMatrixODE})--(\ref{eqConvertedMatrixPDE})
and
(\ref{eqConvertedGarnierODE})--(\ref{eqConvertedGarnierPDE}).
This results in
\begin{eqnarray}
& &
M_{11}={\beta_1 u' + \beta_0 \over e_1},
\\
& &
M_{12}={M_{11,t} - M_{11}^2} - {M_{11} \over (t-u)^2}
- {1 \over t-u} \left({S \over 2} + {3/4 \over (t-u)^2} \right)
\\
& &
M_{21}=t-u,\
\\
& &
L_{21}=- (t-u) C,\
\\
& &
L_{11}= - C M_{11} + {u' - (C + (t-u) C_t) \over 2 (t-u)}
\\
& &
L_{12}={-L_{11,t} + M_{11,x} + L_{21} M_{12} \over t-u}.
\end{eqnarray}

Once expressed in terms of $(t,x,u,u',\abcd)$ only,
this matricial Lax pair has the following features~:
regularity at $t=u$,
concentration of the dependence on $\abcd$ in the elements
 $(L_{12},M_{12})$,
linearity of these two elements in $\abcd$,
absence of any restriction on $\abcd$.
For instance,
the element $L_{11}$ evaluates to 
\begin{eqnarray}
& &
L_{11}={(u-x) \over (x(x-1))},{u-1 \over x},0,0,0
\end{eqnarray}
for the five successive (Pn) of Garnier,
and the element $L_{12}$ to 
\begin{eqnarray}
& &
L_{12}={(\beta_1 u')^2 - \beta_0^2 \over t(t-1)(t-x)u(u-1)},0,0,0,0
\end{eqnarray}
for $a=b=c=d=0$.
The confluence operates on each element separately.

For instance, the matricial Lax pair of (P2') is
\begin{eqnarray}
& &
L= \pmatrix{0 & 0 \cr 1/2 & 0 \cr}
+  \pmatrix{0 & 1 \cr 0   & 0 \cr}
[\beta + 4 \gamma (t+2u) + \delta (t^2 + 2 t u + 3 u^2 +x)]/2,
\\
& &
M= \pmatrix{-u' & 0 \cr t-u & u' \cr}
+  \pmatrix{0 & 1 \cr 0   & 0 \cr}
[2 \alpha + \beta (t+u) + \gamma (4(t^2 + t u + u^2) + 2 x)
\nonumber
\\
& &
\phantom{t^3 + t^2 u + t u^2 + u^3 + u^4 + u^5}
+\delta (t^3 + t^2 u + t u^2 + u^3 + x (t+u))],
\end{eqnarray}
and the one of (J') is
\begin{eqnarray}
& &
L= \pmatrix{0 & 0 \cr 1/2 & 0 \cr}
+  \pmatrix{0 & 1 \cr 0   & 0 \cr}
[\beta + 4 \gamma (\lambda+2 u)+\delta (\lambda^2 + 2 \lambda u + 3 u^2)]/2,
\label{eqLaxJprimeL}
\\
& &
M= \pmatrix{-u' & 0 \cr \lambda-u & u' \cr}
+  \pmatrix{0 & 1 \cr 0   & 0 \cr}
[2 \alpha + \beta (\lambda+u) + 4 \gamma (\lambda^2 + \lambda u + u^2)
\nonumber
\\
& &
\phantom{t^3 + t^2 u + t u^2 + u^3 + u^4 + u^5}
+\delta (\lambda^3 + \lambda^2 u + \lambda u^2 + u^3)],
\label{eqLaxJprimeM}
\end{eqnarray}
in which $\lambda$ is an arbitrary constant.

\section{Rules of discretization of matricial Lax pairs}
\label{sectionDiscretizationRules}
\indent

In order to discretize a given ODE having the Painlev\'e property,
some basic discretization rules have been proposed \cite{CM1996}.

The definition of the (continuous) matricial Lax pair $(L,M)$ can be 
discretized as
\begin{eqnarray}
{\hskip -10.0 truemm}
& &
\psi(x+h/2) = A(x,z) \psi(x-h/2),\
\partial_z \psi(x-h/2) = B(x-h/2,z) \psi(x-h/2),\
\\
{\hskip -10.0 truemm}
& &
\partial_z A(x,z) + A(x,z) B(x-h/2,z) - B(x+h/2,z) A(x,z) =0.
\end{eqnarray}
The continuum limit $h \to 0$ is then
\begin{eqnarray}
& &
{A - 1 \over \pas} \to L,\
(\D z / \D t) B \to M,\
\nonumber
\\
& &
(\D z / \D t) (\partial_z A + A B - \overline{B} A) / \pas
\to 
\partial_t L - \partial_x M + L M - M L,
\label{eqLaxLimiteContinue}
\end{eqnarray}
with some link $F(t,z,\pas)=0$ between the spectral parameters $t$ and $z$.
For a second order equation $E(\overline{u},u,\underline{u},x,\pas)=0$,
the operators $A$ and $B$ must have the $u-$dependences 
$A(\overline{u},u,\underline{u}),B(u,\underline{u})$.

Recently, we proposed \cite{CMMontreal}
a direct method for finding the Lax pair of a discrete equation
whose continuous pair is known.
This consists in discretizing the continuous Lax pair
by obeying the following common sense rules

\begin{enumerate}
\item
conserve the matricial order.
This is indeed the differential order of the scalar Lax pair,
which must be conserved;

\item
replace the continuous spectral parameter $t$ by an unspecified
function $T(z,\pas)$;

\item
discretize the operator $L$ centered at the three points $x-\pas,x,x+\pas$.
If $L$ is traceless,
so is its discretization;

\item
discretize the operator $M$ centered at the two points $x-\pas,x$.
If $M$ is traceless, so is its discretization;


\item
replace each monomial $(\D u / \D x)^k$ by its discretization obeying
the general rules,
multiplied by the $k$-th power of an unspecified function $g(z,\pas)$.
This function $g$,
whose continuum limit must be $1$ for any $z$,
represents the ratio of the stepsize $\pas$ to the differential element $\D x$;
\item
take $B$ as the product of the discretized $M$ by an unspecified function
$J(z,\pas)$
(like Jacobian)
representing a discretization of the derivative $\D T / \D z$;

\item
take $(A-1)/ \pas$ equal to the sum of the discretized operator $L$
and a diagonal matrix of unspecified functions of $(z,\pas)$ only,
diag$(g_1,g_2)$;
these functions,
whose continuum limit must be zero,
account for the dissymmetry of the formula defining $A$.

\end{enumerate}

\section{Application to the discretization of the (Pn) equations
and of their Lax pair}
\label{sectionApplication}
\indent

The (Pn) equations and the pairs of Garnier, 
in their scalar form as well as in their matricial one,
are linear in the four parameters $\abcd$.
If one requires this property to be conserved in the discrete case,
the problem of finding a discrete equation and its discrete Lax pair
splits into five successive subproblems~:
all four parameters zero (or numerical constants),
coefficient of $\alpha$ alone, of $\beta$ alone, etc.

When applied to the autonomous limit (J') of (P2') and its Lax pair
(\ref{eqLaxJprimeL})--(\ref{eqLaxJprimeM}),
the discretization rules give straightforwardly the desired result, namely
\begin{eqnarray}
& &
- {\overline{u} - 2 u + \underline{u} \over h^2}
+ \delta u^2 (\overline{u} + \underline{u})
+ 2 \gamma u (\overline{u} + u + \underline{u})
\nonumber
\\
& &
\phantom{aaaaaaaaaaaaaaaaaaaaaaaaaaaa}
+ \beta
 \left(\lambda_1 u +(1-\lambda_1){\overline{u} + \underline{u} \over 2}\right)
+ \alpha =0,
\\
& &
A/h= \pmatrix{g/h & 0 \cr 1/2 & g/h \cr}
+  \pmatrix{0 & 1 \cr 0   & 0 \cr}
[\beta/2 + 2 \gamma (T+2 u)
\nonumber
\\
& &
\phantom{aaaaaaaaaaaaaaaaaaaaaaaaaaaa}
 +\delta {F(x+h/2) - F(x-h/2) \over u(x+h) - u(x-h)}],
\\
& &
B/J= \pmatrix{-g (u-\underline{u})/h & 0 \cr T - (u + \underline{u})/2 &
               g (u-\underline{u})/h \cr}
+  \pmatrix{0 & 1 \cr 0   & 0 \cr}
[2 \alpha + \beta (T+ (u + \underline{u})/2)
\nonumber
\\
& &
\phantom{t^3 + t^2 u + t u^2 + u^3 + u^4}
 + \gamma (4 T^2 + 2 T (u + \underline{u}) + 4 u \underline{u})
+ \delta F(x-h/2)],
\\
& &
F(x)={T^4 - (u(x+h/2) u(x-h/2))^2 - T^2 (u(x+h/2)-u(x-h/2))^2/2
      \over T - (u(x+h/2) + u(x-h/2))/2},
\\
& &
{g(z)^2 - 1 \over h^2}=
 \beta {2 \lambda_1 -1 \over 4} + \gamma T + \delta {T^2 \over 2},
\\
& &
T'(z)=0.
\end{eqnarray}
In these expressions, $\lambda_1$ is an arbitrary constant
(which could be absorbed in the discrete second derivative),
the functions $J(z)$, $T(z)$ and $g(z)$ are constant,
the spectral parameter is the constant value of $T$.

One notices an unexpected drawback~:
the apparent singularity $t=u$, absent from the continuous matricial pair,
reappears in the coefficient of $\delta$,
because the polynomial $T^3 + T^2 u + T u^2 + u^3$ is discretized as the
quotient of a discretization of $T^4 - u^4$ by
            a discretization of $T   - u  $.

This is the reason why the Lax pair of d--(P2')
obtained by this method is, for the moment, restricted to $\delta=0$
i.e.~to d--(P1)
\begin{eqnarray}
& &
- {\overline{u} - 2 u + \underline{u} \over h^2}
+ \delta (u^2 (\overline{u} + \underline{u}) + x u)
+ \gamma (2 u (\overline{u} + u + \underline{u}) + x)
+ \beta u
+ \alpha =0,
\\
& &
A/h= \pmatrix{g/h & 0 \cr 1/2 & g/h \cr}
+  \pmatrix{0 & 1 \cr 0   & 0 \cr}
[\beta/2 + 2 \gamma (T+2u)],
\\
& &
B/J= \pmatrix{-g (u-\underline{u})/h & 0 \cr T - (u + \underline{u})/2 &
               g (u-\underline{u})/h \cr}
+  \pmatrix{0 & 1 \cr 0   & 0 \cr}
[2 \alpha + \beta (T+ (u + \underline{u})/2),
\nonumber
\\
& &
\phantom{t^3 + t^2 u + t u^2 + u^3 + u^4}
 + \gamma (4 T^2 + 2 T (u + \underline{u}) + 4 u \underline{u} + 2 (x-h/2))],
\\
& &
{g(z)^2 - 1 \over h^2}= \beta/4 + \gamma T,
\\
& &
\gamma T'(z)= \gamma J(z) g(z),
\\
& &
g'(z)=(1/2) \gamma h^2 J(z),
\end{eqnarray}
with $J(z)$ arbitrary.
The contributions $\delta G(x,z)$ and $\delta F(x-h/2,z)$ 
in the elements $A_{12}$ and $B_{12}$
lead to an impossibility.

For (P3'),
the result given in \cite{CMMontreal} deals with the case 
$\alpha=\beta=\gamma=\delta=0$ where the degree of the d--(P3) 
given in Ref.~\cite{GNPRS1994} reduces from two to one
(see Table \ref{tableExistingLax});
in this case a second order discrete Lax pair can be obtained.

\section{Conclusion}
\indent

This direct approach, which serializes the difficulties,
can be improved in the following directions.
If one sticks to the matricial form,
one should design more efficient discretization rules,
so as to at least find the contribution of $\delta$ in the Lax pair of 
d--(P2').
The other possibility is to directly discretize the Garnier pairs on their 
scalar form
(\ref{eqConvertedGarnierODE})--(\ref{eqConvertedGarnierPDE}).

\section*{Acknowledgments}

The financial support of the Tournesol grant T 95/004 is gratefully
acknowledged.
M.~M.~acknowledges the financial support extended within the framework of
the IUAP contract P4/08 funded by the Belgian government.

\vfill \eject


\begin{thebibliography}{99}
\parskip = 5 truept
\small

\bibitem{BZM1987} S.~P.~Burtsev, V.~E.~Zakharov and A.~V.~Mikhailov,
Inverse scattering method with variable spectral parameter,
Teoreticheskaya i Matematicheskaya Fizika {\bf 70} (1987) 323--341
[English: Theor.~and Math.~Phys.~{\bf 70} (1987) 227--240].

\bibitem{ClarksonMansfieldMilne} 
P.~A.~Clarkson, E.~L.~Mansfield and A.~E.~Milne,
Symmetries and exact solutions of a $2+1-$dimensional sine-Gordon system,
Phil.~Trans.~Roy.~Soc.~London A {\bf 354} (1996) 1807--1835.

\bibitem {CM1996} R.~Conte and M.~Musette,
A new method to test discrete Painlev\'e equations,
Phys.~Lett.~A {\bf 223} (1996) 439--438.

\bibitem{CMMontreal} R.~Conte and M.~Musette,
Rules of discretization for Painlev\'e equations,
{\it Theory of nonlinear special functions : the Painlev\'e transcendents}, 
20 pages, 
eds.~L.~Vinet and P.~Winternitz
(Springer, Berlin, 1998).

\bibitem{FlaschkaNewell} H.~Flaschka and A.~C.~Newell,
Monodromy and spectrum preserving deformations, I,
Commun.~Math.~Phys.~{\bf 76} (1980) 65--116.

\bibitem{FokasAblowitz1982} A.~S.~Fokas and M.~J.~Ablowitz,
On a unified approach to transformations and elementary solutions of 
Painlev\'e equations,
JMP {\bf 23} (1982) 2033--2042.

\bibitem{FGR} A.~S.~Fokas, B.~Grammaticos and A.~Ramani,
From continuous to discrete Painlev\'e equations,
Journal of mathematical analysis and applications {\bf 180} (1993) 342--360.

\bibitem{Fuchs1905} R.~Fuchs,
Sur quelques \'equations diff\'erentielles lin\'eaires du second ordre,
\CRAS\ {\bf 141} (1905) 555--558. 

\bibitem{Fuchs1907} R.~Fuchs,
\"Uber lineare homogene Differentialgleichungen zweiter Ordnung mit drei im
Endlichen gelegenen wesentlich singul\"aren Stellen,
Math.~Annalen {\bf 63} (1907) 301--321.
 
\bibitem {GambierThese} B.~Gambier,
Sur les \'equations diff\'erentielles du second ordre et du premier degr\'e
dont l'int\'egrale g\'en\'erale est \`a points critiques fixes,
Th\`ese, Paris (1909); Acta Math.~{\bf 33} (1910) 1--55.
 
\bibitem{GarnierThese} R.~Garnier,                                    
Sur des \'equations diff\'erentielles du troisi\`eme ordre dont
l'int\'e\-gra\-le
g\'en\'erale est uniforme et sur une classe d'\'equations nouvelles d'ordre
sup\'e\-ri\-eur dont l'int\'egrale g\'en\'erale a ses points critiques fixes,
Th\`ese, Paris (1911); 
\AnnENS\ {\bf 29} (1912) 1--126.

\bibitem{GNPRS1994} B.~Grammaticos, F.~W.~Nijhoff, V.~Papageorgiou, A.~Ramani
 and J.~Satsuma,
Linearization and solutions of the discrete Painlev\'e III equation,
Phys.~Lett.~A {\bf 185} (1994) 446--452.

\bibitem{GNRCargese}
B.~Grammaticos, F.~W.~Nijhoff and A.~Ramani,
Discrete Painlev\'e equations,
{\it The Painlev\'e property, one century later}, 
114 pages,
ed.~R.~Conte,
CRM series in mathematical physics (Springer, Berlin, 1998).

\bibitem{Gromak1975} V.~I.~Gromak,
Theory of Painlev\'e's equation,
Differentsial'nye Uravneniya {\bf 11} (1975) 373--376
[English: Diff.~equ.~{\bf 11} (1975) 285--287].

\bibitem {JimboMiwaII} M.~Jimbo and T.~Miwa,
Monodromy preserving deformations of linear ordinary differential equations
with rational coefficients. II,
Physica D {\bf 2} (1981) 407--448.
 
\bibitem{JS1996} M.~Jimbo and H.~Sakai,
A $q-$analog of the sixth Painlev\'e equation,
Lett.~Math.~Phys.~{\bf 38} (1996) 145--154.

\bibitem{KRT} A.~V.~Kitaev, A.~V.~Rybin and J.~Timonen,
Similarity solutions of the deformed Maxwell-Bloch system,
J.~Phys.~A {\bf 26} (1993) 3583--3595.

\bibitem{CargeseMahoux} G.~Mahoux,
Introduction to the theory of isomonodromic deformations
of linear ordinary differential equations with rational coefficients,
{\it The Painlev\'e property, one century later}, 
44 pages,
ed.~R.~Conte,
CRM series in mathematical physics (Springer, Berlin, 1998).

\bibitem{MilneThesis} A.~Milne,
Ph.~D.~Thesis (University of Exeter, UK, 1995).

\bibitem {PaiCRAS1906}
P.~Painlev\'e,
Sur les \'equations diff\'erentielles du second ordre \`a points critiques
fixes,
\CRAS\ {\bf 143} (1906) 1111--1117.
 
\bibitem {PNGR1992} V.~G.~Papageorgiou, F.~W.~Nijhoff, B.~Grammaticos
 and A.~Ramani,
Isomonodromic deformation problems for discrete analogs of \Plv\ equations,
Phys.~Lett.~A {\bf 164} (1992) 57--64.

\bibitem{Poincare1883} H.~Poincar\'e,
Sur les groupes des \'equations lin\'eaires,
Acta mathematica {\bf 4} (1883) 201--312.

\bibitem {Potts1987} R.~B.~Potts (1987),
Weierstrass elliptic difference equations,
Bull.~Austral.~Math.~Soc.~{\bf 35} (1987) 43--48.

\bibitem {RGH1991} A.~Ramani, B.~Grammaticos and J.~Hietarinta,        
Discrete versions of the Painlev\'e equations,
Phys.~Rev.~Lett.~{\bf 67} (1991) 1829--1832.

\bibitem{RGO1998} A.~Ramani, B.~Grammaticos and Y.~Ohta,
The Painlev\'e of discrete equations and other stories,
{\it Theory of nonlinear special functions : the Painlev\'e transcendents},
eds.~L.~Vinet and P.~Winternitz (Springer, Berlin, 1998).

\bibitem{Heun1889} 
K.~Heun,
Zur Theorie der Riemann'schen Funktionen zweiter Ordnung mit vier Verzweigungspunkten,
{bf 33} (1889) 161--179.

\bibitem{HeunWorkshop} 
A.~Seeger and W.~Lay (eds.),
{\it Centennial workshop on Heun's equation -- Theory and applications}
(Max-Planck-Institut, Stuttgart, 1990).

\bibitem{WinternitzSainteAdele} P.~Winternitz,
Physical applications of Painlev\'e type equations quadratic in the highest
derivatives,
{\it Painlev\'e transcendents, their asymptotics and physical applications},
425--431,
eds.~D.~Levi and P.~Winternitz (Plenum, New York, 1992).

\end{thebibliography}
\end{document}